\documentclass[UTF8,a4paper]{article}
\usepackage{amsmath}
\usepackage{graphicx}
\usepackage{subfigure}
\usepackage{epstopdf}
\usepackage{geometry}
\usepackage{ulem}
\usepackage{indentfirst}
\usepackage{amsfonts,amssymb,dsfont}
\usepackage{setspace}
\usepackage{threeparttable}
\usepackage{amsmath,mathtools,amsthm}
\usepackage{array}
\usepackage{extarrows}
\usepackage{upgreek}

\makeatletter
\renewcommand*\env@matrix[1][\arraystretch]{%
  \edef\arraystretch{#1}%
  \hskip -\arraycolsep
  \let\@ifnextchar\new@ifnextchar
  \array{*\c@MaxMatrixCols c}}
\makeatother
\begin{document}

%QED 3+1: Appelquist T W, Bowick M, Karabali D and Wijewardhana L C R 1986 Phys. Rev. D 33 3704
%Optical evidence for a Weyl semimetal state in pyrochlore Eu 2 Ir 2 O 7
%Collective Modes of the Massless Dirac Plasma
%relaxation time:Dc and ac transport in silicene
%Femtosecond carrier dynamics and saturable absorption in graphene

\title{\bf Dynamical current-current correlation in the two-dimensional parabolic Dirac system}
\author{Chen-Huan Wu
\thanks{chenhuanwu1@gmail.com}
%\\Key Laboratory of Atomic $\&$ Molecular Physics and Functional Materials of Gansu Province,
\\College of Physics and Electronic Engineering, Northwest Normal University, Lanzhou 730070, China}

\maketitle
\vspace{-30pt}
\begin{abstract}
\begin{large} 

We theoretically investigate the current-current correlation of the two-dimensional (2D) parabolic Dirac system
in hexogonal lattice.
The analytical expressions of the random phase approximation (RPA) susceptibility,
Ruderman-Kittel-Kasuya-Yosida (RKKY) Hamiltonian, and the diamagnetic orbital susceptibility in noninteracting case
base on the density-density or current-current correlation function are derived
and quantitatively analyzed.
In noninteracting case, the dynamical polarization within RPA and spin transverse susceptibility 
as well as the RKKY interaction (when close to the half-filling)
are related to the the current-current response in the 2D parabolic Dirac system.
Both the case of anisotropic dispersion and isotropic dispersion are discussed.
\\
{\bf PACS number(s)}: 73.20.At, 67.85.De\\

\end{large}

\end{abstract}
\begin{large}

\section{Introduction}

For the extrinsic parabolic 2D parabolic hexagonal Dirac system with finite chemical potential (larger than the band gap here)
and away from half-filling,
the luttinger liquid model emerges in the absence of the umklapp scattering,
due to this reason, the singularity at the nesting wave vector $2k_{F}$ is missing,
and thus the Friedel oscillation at $2k_{F}$ is also vanishes.
%{RKKY interaction and Kondo screening cloud for strongly correlated electrons}
That's in contrast to the case of the perfect nesting Fermi surface at half-filling,
where the divergence of the real spin susceptibility is obvious at the nesting wave vector. 
For the usual case at (or near) half-filling, the RKKY interaction has a magnitude oscillation at nesting wave vector
%\cite{}
due to the non-analytical susceptibility,
but this phenomenon vanishes when away from the half-filling.
Furthermore, for the helical Luttinger liquid,
which can be realized by, e.g., the quantum spin Hall edge state\cite{Ezawa M},
%{Monolayer Topological Insulators Silicene, Germanene, and Stanene}
the spin rotation during the back scattering would eliminates the oscillation at nesting wave vector\cite{Yevtushenko O M}.
and the kondo screening is dominates over the RKKY interaction at large distance (between magnetic impurities) for such special Luttinger liquid.

We focus on the equilibrium dynamics of the extrinsic 2D parabolic Dirac system with a finite chemical potential 
in this letter.
The current-current correlation as well as the related anomalous divergence of the diamagnetic susceptibility 
in the presence of a staic weak magnetic field is
studied.
The magnitude oscillations at the nesting wave vector between two Dirac nodes are not being discussed too much in this letter,
we focus on the response functions as well as the important integrals with the ultraviolet cutoff.
The analytical results about the momentum space Green's function and spectral function are presented,
which are important to the study of the optical conductivity and the plasmon dispersion.
%{Dynamic current-current susceptibility in three-dimensional Dirac and Weyl semimetals}
Our results are also significative to the study of the dynamical susceptibility or RKKY interaction
in 1D Luttinger liquid model
and the 3D Dirac or Weyl systems where the longitudinal response function are also needed.

\section{RPA susceptibility and spectral function}

It's well known that the random-phase-approximation (RPA) is available for the doped Dirac system
where the chemical potential is larger than the temperature spacing.
In this case, we can successfully recover the Fermi liquid picture for the Dirac system,
where the density of states (DOS) at Fermi level $D_{F}$ has the following semiclassical relation
%{Electronic transport in graphene A semiclassical approach including midgap states}
\begin{equation} 
\begin{aligned}
D_{F}=\frac{g}{S}\sum_{k}\frac{\partial f_{k}}{\partial \varepsilon},
\end{aligned}
\end{equation}
where $g=g_{s}g_{v}=4$ denotes the spin and valley degrees of freedom, 
and $f_{k}$ is the electron distribution function within the Boltzmann theory,
and $S$ is the area of the unit cell.
Since the magnetic coupling between magnetic impurities via the double-exchange mechanism can be mediated by the conduction electrons,
the localized spin interactions give rise to the ferromagnetism when the magnetic impurities are close to each other
and overwhelm the conduction electrons,
%{Localized spin ordering in Kondo lattice models}
that's competes with the Kondo coupling as well as the fluctuation of the spin (or pseudospin) singlets.
In order to describe the screening of the Fermi liquid picture,
we explore the non-static RPA susceptibility
\begin{equation} 
\begin{aligned}
\Pi_{00}(\omega,q)=-\frac{g}{S}\int^{\frac{1}{T}}_{0} d\tau e^{i\omega_{m}\tau}\langle T_{\tau}J_{0}(q,\tau)J_{0}(-q,0)\rangle,
\end{aligned}
\end{equation}
where $J_{0}=-ite\sum_{n,\langle i,j\rangle}(c_{in}^{\dag}c_{jn}+H.c.)$ is the paramagnetic current operator
with $n$ the layer index.
In order to make the current-current correlation function obey the gauge invariance,
the Peierls substitution can be applied.
%{Exact renormalization group computation of the optical conductivity of graphene}
$\omega_{m}=2\pi m T(m=0,\pm 1,\pm 2\cdot\cdot\cdot)$ is the Bosonic Matsubara frequency, 
$T_{\tau}$ is the imaginary time-order operator,
and $\langle\cdot\cdot\cdot\rangle$ denotes the imaginary-time-average over the 
whole canonical ensemble.
To first order of the electron interaction,
we obtain the intraband ($s=1$) or interband ($s=-1$) polarization through the density-density correlation function
which is
\begin{equation} 
\begin{aligned}
\chi^{00}(\omega,k)=\frac{T}{S}\sum_{k,\omega}{\rm Tr}[I_{4\times 4}G(\Omega,k)I_{4\times 4}G(\omega',k')],
\end{aligned}
\end{equation}
where $k'=k+q$, $\omega'=\omega+\Omega$, 
$\Omega_{n}=2\pi (n+1) T(n=0,\pm 1,\pm 2\cdot\cdot\cdot)$ is the Fermionic Matsubara frequency.
Here we take the replacement $i\omega_{m}=\omega+i\eta$ and $i\Omega_{m}=\Omega+i\eta$ 
through the real line version of the Sokhotski-Plemelj theorem, where the $\omega$ can be replaced by the renormalized dispersion
$\omega=\frac{1}{g}\sum_{s\tau}(\varepsilon^{s\tau}+\varepsilon^{-s-\tau})$.
After unitary transformation, the Green's function in spin basis reads
\begin{equation} 
\begin{aligned}
G(\omega,k)=\frac{I_{4\times 4}}{\omega^{2}-\varepsilon^{2}}
\begin{pmatrix}[1.5]
g_{\uparrow} &0\\
0&g_{\downarrow}
\end{pmatrix},
\end{aligned}
\end{equation}
where the element $g$ is
\begin{equation} 
\begin{aligned}
g_{\uparrow(\downarrow)}=
\begin{pmatrix}[1.5]
\omega^{++(-+)} &k_{-}\\
k_{+}&\omega^{+-(--)}
\end{pmatrix},
\end{aligned}
\end{equation}
here the indices of $\omega$ denote the spin and pseudospin indices, respectively,
and $k_{\mp}=k_{x}\mp ik_{y}=ke^{\mp i\theta_{k}}$ where $\theta_{k}$ is dependent on the direction of $k$.
By substituting the Green's function into Eq.(3), the density-density correlation function can be rewritten as
\begin{equation} 
\begin{aligned}
\chi^{00}(\omega,k)=\frac{T}{S}\sum_{k,\omega}[2k_{+}k'_{-}+2k_{-}k'_{+}+\sum_{s\tau}\omega^{s\tau}\omega'^{s\tau}].
\end{aligned}
\end{equation}
Then the polarization for extrinsic model can be obtained as
\begin{equation} 
\begin{aligned}
\Pi^{\pm}(\omega,q)=-\frac{gqV_{q}^{b}}{S}\int\frac{dk^{2}}{(2\pi)^{2}}
\frac{1}{2}(1\pm{\rm cos}2\theta)
(\frac{n_{F}(\varepsilon_{k})}{\omega+i\eta+\varepsilon_{k}\mp\varepsilon_{k'}}
-\frac{n_{F}(\varepsilon_{k'})}{\omega+i\eta-\varepsilon_{k}\pm\varepsilon_{k'}})
\end{aligned}
\end{equation}
where the superscript $\pm$ correspond to the intraband and interband transition, respectively,
and the bare Coulomb interaction is $V_{q}^{b}=2\pi e^{2}e^{-qd}/\varepsilon_{0}\varepsilon q$
with $d$ the distance from the 2D sheet.
$\varepsilon_{k}$ is the eigenerengy.
${\rm cos}\theta=\frac{kk'{\rm cos}\phi+D^{2}}{\varepsilon_{k}\varepsilon_{k'}}$
where $\phi$ is the angle between $k$ and $q$.
The Dirac-Fermi distribution function $n_{F}$ at low-temperature can be estimated as 
$n_{F}(\varepsilon_{k})=\theta(\mu-\varepsilon_{k})$ while at high temperature it can be estimated as $n_{F}=1/2$.
For simplicity, we only discuss the low-temperature case.
The low-energy effective Hamiltonian of the parabolic 2D Dirac system reads
\begin{equation} 
\begin{aligned}
H=\eta(Ds_{z}\tau_{z}-\frac{({\bf k}\cdot{\pmb \tau})^{2}}{2m})-\mu,
\end{aligned}
\end{equation}
where $\eta=\pm 1$ denotes the valley degree of freedom, $s_{z}$ and ${\pmb \tau}$ denote the spin and pseudospin degrees of freedom, respectively.
$D$ is the Dirac mass, $\mu$ is the chemical potential.
The eigenenergies can be obtained through solving the above Hamiltonian:
\begin{equation} 
\begin{aligned}
\varepsilon_{\pm}=\frac{-2m\mu\pm \sqrt{k^{4}+4m^{2}(D^{2}+\mu^{2})+4\eta k^{2}m\mu{\rm cos}2\Phi}}{2m},
\end{aligned}
\end{equation}
where $\phi={\rm atan}k_{y}/k_{x}$ and isotropic dispersion corresponds to the
$\Phi=\pi/4$.
However, we firstly focus on the extremely anisotropic case with $k_{y}$ much larger than the $k_{x}$
and with $\Phi=\pi/2$.

Through the real line version of the Sokhotski-Plemelj theorem with the retarded Green's function,,
the imaginary part of the extrinsic polarization reads
\begin{equation} 
\begin{aligned}
{\rm Im}\ \Pi^{\pm}(\omega,q)=-\frac{g V_{q}^{b}}{4\pi^{2}S}\int^{2\pi}_{0}d\phi\int^{\Lambda}_{0}dk k
\frac{1}{2}(1\pm{\rm cos}2\theta)
(\delta(\omega-\varepsilon_{k}\pm\varepsilon_{k'})-\delta(\omega+\varepsilon_{k}\mp\varepsilon_{k'})).
\end{aligned}
\end{equation}
For $\omega=\varepsilon_{k}\mp\varepsilon_{k'}$,
the above expression can be written as
\begin{equation} 
\begin{aligned}
{\rm Im}\ \Pi^{\pm}(\omega,q)=-\frac{g V_{q}^{b}}{4\pi^{2}S}\int^{\Lambda}_{0}dk k
\left(\pi\mp\frac{\pi(-k^{2}k'^{2}-2D^{4}+\varepsilon_{k}^{2}\varepsilon_{k'}^{2})}{\varepsilon_{k}^{2}\varepsilon_{k'}^{2}}\right),
\end{aligned}
\end{equation}
after integrate over the angle $\phi$,
and the real part can be obtained by using the Kramers-Kronig relation.
%\begin{equation} 
%\begin{aligned}
%{\rm Re}\ \Pi^{\pm}(\omega',q)=\frac{2}{\pi}P.V.\int^{\Lambda}_{0}\frac{d\omega}{\omega-\omega'}{\rm Im}\Pi^{\pm}(\omega,q),
%\end{aligned}
%\end{equation}
The $\Lambda$ is the Fourier transform of the ultraviolet cutoff in order to carry out the regularization,
and it's needed here for 2D Dirac systems (like the graphene or silicene) 
since the momentum integration won't be ultraviolet convergent unlike the QED\cite{Kotov V N}.
In extrinsic case, for calculation,
we set the parameters as $m=0.268 m_{0}$, $D=0.02$ eV, $\mu=0.2$ eV,
then for the ratio $\alpha=\varepsilon_{k'}/\varepsilon_{k}$
can be approximated obtained as
$\alpha=\frac{1.86}{k^{2}}(q+k)^{2}$ (in expanded form),
and thus we have $\varepsilon_{k}\varepsilon_{k'}\approx (6.48k^{2}+1.315)(k+q)^{2}$.
We further set the momentum cutoff equals to the bandwidth $W=3t=4.8$ eV,
then the imaginary part of the polarization can be evaluated analytically
as shown in Fig.1,
where we only show the part of momentum which excess the Fermi surface.
The logarithmic screening of the Dirac Fermion can be seen in the Fig.1,
and that's in contrast to the case of strong screening in Thomas-Fermi approximation.
%{Quantum criticality of topological phase transitions in three-dimensional interacting electronic systems}
It's worth to note that, at zero-temperature,
the polarization in one-loop configuration (the two-point density-density correlation) is periodic and thus
it won't relaxes into the steady state\cite{Cazalilla M A,Wu C Htime}.
Since the current-current correlation doesn't has any more time-dependent terms compared to the density-density correlation,
we can conclude that the current-current correlation at zero-temperature in one-loop order.
%{One loop integrals reduction}

To present the spectroscopic features during the transition,
we use the spectral function.
Base on the retarded Green's function,
the spectral function reads
\begin{equation} 
\begin{aligned}
A(\omega,q)
=&\frac{-1}{\pi}{\rm Im}\int\int dtdre^{i\omega t-iqr}G(r,t)\\
=&\frac{-1}{\pi}{\rm Im}G(\omega,q)\\
=&\delta(\omega-\varepsilon).
\end{aligned}
\end{equation}
In spin basis, the above expression can be rewritten as
\begin{equation} 
\begin{aligned}
A(\omega,q)=
\begin{pmatrix}[1.5]
A_{\uparrow} &0\\
0&A_{\downarrow}
\end{pmatrix},
\end{aligned}
\end{equation}
where
\begin{equation} 
\begin{aligned}
A_{\uparrow}=\frac{1}{2|{\bf k}|}
\begin{pmatrix}[1.5]
\delta(\omega-\varepsilon^{++}) &k_{-}\delta(\omega-\varepsilon^{++})\\
k_{+}\delta(\omega-\varepsilon^{-+})&\delta(\omega-\varepsilon^{+-})\\
\end{pmatrix},
A_{\downarrow}=\frac{1}{2|{\bf k}|}
\begin{pmatrix}[1.5]
\delta(\omega-\varepsilon^{-+}) &-k_{-}\delta(\omega-\varepsilon^{--})\\
-k_{+}\delta(\omega-\varepsilon^{-+})&\delta(\omega-\varepsilon^{--})\\
\end{pmatrix}.
\end{aligned}
\end{equation}
The retarded Green's function before Fourier transform reads
\begin{equation} 
\begin{aligned}
G_{ij}(t-t')=&-i\theta(t-t')(\{c_{i}(t),c_{j}^{\dag}(t')\})\\
=&i\theta(t-t')(\int^{\infty}_{0}A(\omega,q)n_{F}(\omega)e^{i\omega(t-t')}d\omega
+\int^{\infty}_{0}A(\omega,q)(1-n_{F}(\omega))e^{i\omega(t-t')}d\omega),
\end{aligned}
\end{equation}
where $A(\omega,q)n_{F}$ describes the occupied spectrum,
$c$ is the Grassmann field.
That implies that for equilibrium case with translational invariance, 
the system is dominated by the spectral function which contains the information of the band dispersion and quasiparticle lifetime,
%{Correlated Electrons out of Equilibrium Non-Equilibrium DMFT}
thus,
within the fluctuation-dissipation theorem at finite temperature,
the imaginary part of the response function (i.e., the retarded Green's function here) is related to the power spectrum of the fluctuation
field.
The assumpation of the translational invariance is also important to the real space spin susceptibility\cite{Stano P}.
Note that 
for nonequilibrium case,
the distribution function in above expression is no longer the Dirac-Fermi type.

\section{RKKY interaction}

When the distance between two magnetic impurities is larger than the ultraviolet cutoff
$\Lambda_{r}=\frac{1}{2\pi}\int^{\infty}_{0}\Lambda e^{i{\bf k}\cdot{\bf r}}dk$,
the ferromagnetic phase vanishes due to the vanishing magnetic coupling which is via the double-exchange mechanis,
%{Magnetism in the dilute Kondo lattice model}
thus then the RKKY interaction is affected more by the other phases in the Bose-Hubbard model,
especially when the next-nearest-neighbor (NNN) complex hopping is taken into account\cite{Nakafuji T,Peres N M R}.
The phase factor with the NNN hopping is possible experimentally by the time-periodic driving field in optical lattice
which is also a great platform to exploring the quenching dynamics\cite{Cazalilla M A} where the Wigner-Seitz radius can also be controlled.
The RKKY interaction in second-order perturbation reads
\begin{equation} 
\begin{aligned}
H_{RKKY}=J^{2}\sum_{\alpha\beta}{\bf I}_{1\alpha}\chi_{\alpha\beta}{\bf I}_{2\beta},
\end{aligned}
\end{equation}
where ${\bf I}$ is the magnetic moment of the magnetic impurities,
$J$ is the spin exchange interaction (or $s-d$ exchange interaction).
The spin susceptibility $\chi_{\alpha\beta}$ is related to the occupied spectrum $A(\omega,q)n_{F}(\omega)$
which will be discussed later.

We only consider the spin splitting (the two-band model) here
and ignore the Rashba effect as well as the trigonal warping (i.e., the $k^{3}$-term).
The retarded Green's function in real space can be obtained through the Fourier transform of that in the momentum space
which we write here in diagonal basis as
\begin{equation} 
\begin{aligned}
G(\omega,k)=
\begin{pmatrix}[1.5]
\frac{1}{\omega-\varepsilon_{+}} &0\\
0&\frac{1}{\omega-\varepsilon_{-}}\\
\end{pmatrix},
\end{aligned}
\end{equation}
where $I_{0}(x)$ is the modified Bessel function of the first kind at zeroth order.
and then the real space one can be written as
\begin{equation} 
\begin{aligned}
G_{\pm}(\omega,\pm r)=&\int\frac{d^{2}k}{(2\pi)^{2}}e^{\pm i{\bf k}\cdot{\bf r}}\frac{1}{\omega-\varepsilon_{\pm}},\\
=&\int^{\pi}_{0}d\theta\int^{\Lambda}_{0}dk ke^{\pm i{\bf k}\cdot{\bf r}}\frac{1}{\omega-\varepsilon_{\pm}},\\
=&\pi I_{0}(\pm ikr) \int^{\Lambda}_{0}dk k\frac{1}{\omega-\varepsilon_{\pm}},\\
=&\pi I_{0}(\pm ikr)
\begin{pmatrix}[1.5]
\int^{\Lambda}_{0}\frac{kdk}{\omega-\varepsilon_{+}} &0\\
0&\int^{\Lambda}_{0}\frac{kdk}{\omega-\varepsilon_{-}}\\
\end{pmatrix},
\end{aligned}
\end{equation}
The analytical evaluation of the real space Green's function is easy to obtained by integral over the momentum with a determined ultraviolet cutoff
(setted as equals to the bandwidth here for the two-band model here).
We follow the parameters defined in above,
then the eigenenergies for the extrinsic case can be estimated as
\begin{equation} 
\begin{aligned}
\varepsilon_{\pm}=\frac{\pm\sqrt{k^{4}+0.0116-0.2144\eta k^{2}}}{0.536}-0.2,
\end{aligned}
\end{equation}
then the diagonal elements within the above expression of the real space Green's function in static limit can be obtained as shown
in the Fig.2.
From Fig.2, we show the diagonal elements of $G_{\pm}(\omega,\pm r)$ in spin splitted two-band model
as a function of the ultraviolet cutoff.
Here we only consider the static limit with $\omega\rightarrow 0$
since the $\omega$ here brings large numberical fluctuation during the integral process over the momentum $k$.
In such limit, after some tedious but straightforward computations,
we have 
\begin{equation} 
\begin{aligned}
\lim_{\omega\rightarrow 0}\int^{\Lambda}_{0}\frac{kdk}{\omega-\varepsilon_{\pm}}=&\pm 0.268{\rm asinh}(10.3077-96.1538\Lambda^{2})\bigg|^{\Lambda}_{0}\\
=&\pm 0.268{\rm asinh}(10.3077-96.1538\Lambda^{2})\mp 0.8116,
\end{aligned}
\end{equation}
which is the result as presented in Fig.2.

By summing up all the eigenstates of the real space Green's function,
the RKKY interaction can be devided into three terms,
Heisenberg term, Ising term, and Dzyaloshinskii-Moriya (DM) term in the absence of the inversion symmetry,
which can be distinced by the spin susceptibility tensor.
The spin susceptibility tensors have a similar form to the occupied spectral function,
which read\cite{Wu C Hrkky}
\begin{equation} 
\begin{aligned}
\chi_{H}=&-\frac{2}{\pi}{\rm Im}\int^{\mu}_{-\infty}d\omega{\rm Tr}[\sigma_{x}G_{\pm}(\omega, r)\sigma_{x}G_{\pm}(\omega,-r)],\\
\chi_{I}=&-\frac{2}{\pi}{\rm Im}\int^{\mu}_{-\infty}d\omega({\rm Tr}[\sigma_{z}G_{\pm}(\omega, r)\sigma_{z}G_{\pm}(\omega,-r)]
-{\rm Tr}[\sigma_{x}G_{\pm}(\omega, r)\sigma_{x}G_{\pm}(\omega,-r)]),\\
\chi_{DM}=&-\frac{2}{\pi}{\rm Im}\int^{\mu}_{-\infty}d\omega{\rm Tr}[\sigma_{x}G_{\pm}(\omega, r)\sigma_{y}G_{\pm}(\omega,-r)].
\end{aligned}
\end{equation}
By substituting the above expression of real space Green's function into the spin susceptibility tensors,
we obtain
\begin{equation} 
\begin{aligned}
\chi_{H}=&-\frac{2}{\pi}{\rm Im}\int^{\mu}_{-\infty}d\omega\\
&\left(\pi I_{0}(ikr)\int^{\Lambda}_{0}\frac{kdk}{\omega-\varepsilon_{-}}\pi I_{0}(-ikr)\int^{\Lambda}_{0}\frac{kdk}{\omega-\varepsilon_{+}}
+\pi I_{0}(ikr)\int^{\Lambda}_{0}\frac{kdk}{\omega-\varepsilon_{+}}\pi I_{0}(-ikr)\int^{\Lambda}_{0}\frac{kdk}{\omega-\varepsilon_{-}}\right),\\
\chi_{I}=&-\frac{2}{\pi}{\rm Im}\int^{\mu}_{-\infty}d\omega\\
&\left[
\left(\pi I_{0}(ikr)\int^{\Lambda}_{0}\frac{kdk}{\omega-\varepsilon_{+}}\pi I_{0}(-ikr)\int^{\Lambda}_{0}\frac{kdk}{\omega-\varepsilon_{+}}
+\pi I_{0}(ikr)\int^{\Lambda}_{0}\frac{kdk}{\omega-\varepsilon_{-}}\pi I_{0}(-ikr)\int^{\Lambda}_{0}\frac{kdk}{\omega-\varepsilon_{-}}\right)\right.\\
&\left.-\left(\pi I_{0}(ikr)\int^{\Lambda}_{0}\frac{kdk}{\omega-\varepsilon_{-}}\pi I_{0}(-ikr)\int^{\Lambda}_{0}\frac{kdk}{\omega-\varepsilon_{+}}
+\pi I_{0}(ikr)\int^{\Lambda}_{0}\frac{kdk}{\omega-\varepsilon_{+}}\pi I_{0}(-ikr)\int^{\Lambda}_{0}\frac{kdk}{\omega-\varepsilon_{-}}\right)
\right],\\
\chi_{DM}=&-\frac{2}{\pi}{\rm Im}\int^{\mu}_{-\infty}d\omega\\
&i\left(\pi I_{0}(ikr)\int^{\Lambda}_{0}\frac{kdk}{\omega-\varepsilon_{-}}\pi I_{0}(-ikr)\int^{\Lambda}_{0}\frac{kdk}{\omega-\varepsilon_{+}}
-\pi I_{0}(ikr)\int^{\Lambda}_{0}\frac{kdk}{\omega-\varepsilon_{+}}\pi I_{0}(-ikr)\int^{\Lambda}_{0}\frac{kdk}{\omega-\varepsilon_{-}}\right)
.
\end{aligned}
\end{equation}
Then the RKKY Hamiltonian can be obtained as
\begin{equation} 
\begin{aligned}
H_{RKKY}=J^{2}[\chi_{H}{\bf I}_{1}\cdot{\bf I}_{2}+\chi_{I}I_{1}^{z}I_{2}^{z}+\chi_{DM}({\bf I}_{1}\times{\bf I}_{2})_{z}],
\end{aligned}
\end{equation}
constituted by the integral $\int^{\Lambda}_{0}\frac{kdk}{\omega-\varepsilon_{\pm}}$ as shown in Fig.2 for anisotropic dispersion.

\section{Orbital susceptibility}

In the presence of a weak static magnetic field,
the diamagnetic orbital susceptibility reads\cite{Koshino M,Fukuyama H}
\begin{equation} 
\begin{aligned}
\chi_{{\rm orb}}(q)=-\frac{e^{2}T\hbar^{2}v_{F}^{2}}{6\pi c^{2}}\lim_{q\rightarrow 0}\frac{\Pi_{xy}(q)}{q^{2}},
\end{aligned}
\end{equation}
in the optical limit (long-wavelength limit),
which is given by the relation
\begin{equation} 
\begin{aligned}
\chi_{{\rm orb}}(q)=\lim_{B\rightarrow 0}\frac{M}{B},
\end{aligned}
\end{equation}
where $M=-\frac{\partial \mathfrak{E}}{\partial B}$ is the magnetization
and $\mathfrak{E}$ here is the groud state energy at zero-temperature or the thermodynamical potential 
%{Magnetic torque anomaly in the quantum limit of Weyl semimetals}
at finite temperature\cite{Fukuyama H}.
The transverse susceptibility has $\Pi_{xy}(q)=\Pi_{xx}(q)\rightarrow 0$ according to diamagnetic sum rule.
The transverse susceptibility $\Pi_{xy}(q)$ can be obtained by the transverse curren-current correlation function
\begin{equation} 
\begin{aligned}
\chi^{xy}(\omega,q)=&\frac{T}{S}\sum_{k,\omega}{\rm Tr}[\sigma_{x}G(\Omega,k)\sigma_{y}G(\omega',k')]\\
=&\frac{T}{S}\sum_{k,\omega}
i(\frac{1}{\Omega-\varepsilon_{-}}\frac{1}{\omega'-\varepsilon'_{+}}-\frac{1}{\Omega-\varepsilon_{+}}\frac{1}{\omega'-\varepsilon'_{-}}),
\end{aligned}
\end{equation}
where $\omega'=\omega+q$ as defined above
and $\varepsilon'$ is related to the momentum $(k+q)$,
i.e.,
\begin{equation} 
\begin{aligned}
\varepsilon'_{\pm}=\frac{-2m\mu\pm \sqrt{(k+q)^{4}+4m^{2}(D^{2}+\mu^{2})+4\eta (k+q)^{2}m\mu{\rm cos}2\Phi}}{2m}.
\end{aligned}
\end{equation}

The transverse spin susceptibility can be written as
\begin{equation} 
\begin{aligned}
\Pi_{xy}(q)=&gV^{b}_{q}\chi^{xy}\\
=&-gV^{b}_{q}\frac{T}{S}\sum_{k,\omega}
\lim_{q\rightarrow 0}
i(\frac{1}{\Omega-\varepsilon_{-}}\frac{1}{\omega'-\varepsilon'_{+}}-\frac{1}{\Omega-\varepsilon_{+}}\frac{1}{\omega'-\varepsilon'_{-}}).
\end{aligned}
\end{equation}
To proceed further,
we convert the summation over $k$ into the interal in  mometum space,
then the transverse spin susceptibility becomes
\begin{equation} 
\begin{aligned}
\Pi_{xy}(q)=&gV^{b}_{q}\chi^{xy}\\
=&gV^{b}_{q}\frac{T}{S}\sum_{\omega}\int^{\pi}_{0}d\theta\int^{\Lambda}_{0}\frac{kdk}{(2\pi)^{2}}
\lim_{q\rightarrow 0}
i(\frac{1}{\Omega-\varepsilon_{-}}\frac{1}{\omega'-\varepsilon'_{+}}-\frac{1}{\Omega-\varepsilon_{+}}\frac{1}{\omega'-\varepsilon'_{-}}).
\end{aligned}
\end{equation}
According to usual dealing way to the orbital susceptibility\cite{Fukuyama H,Giuliani G,Thakur A}, 
we still focus on the static case here (with the weak static magnetic field).

Since at $q=0$, we have
\begin{equation} 
\begin{aligned}
\frac{T}{S}\sum_{k,\omega}
i(\frac{1}{\Omega-\varepsilon_{-}}\frac{1}{\omega'-\varepsilon'_{+}}-\frac{1}{\Omega-\varepsilon_{+}}\frac{1}{\omega'-\varepsilon'_{-}})
=
\frac{T}{S}\sum_{k,\omega}
i(\frac{1}{\Omega-\varepsilon_{-}}\frac{1}{\omega'-\varepsilon_{+}}-\frac{1}{\Omega-\varepsilon_{+}}\frac{1}{\omega'-\varepsilon_{-}}),
\end{aligned}
\end{equation}
the static transverse susceptibility vanishes in such case,
thus we know that the finite $q$ is important to the diamagnetic susceptibility,
For $q=0$ case, after some tedious but straightforward computations,
we obtain the analytical expression in staic limit
\begin{equation} 
\begin{aligned}
\mathfrak{R}
=&\int^{\Lambda}_{0}\frac{kdk}{\varepsilon_{-}\varepsilon_{+}}\\
=&-0.0025{\rm atan}(536-5000k^{2})\bigg|^{\Lambda}_{0}\\
=&-0.0025{\rm atan}(536-5000k^{2})+0.003948,
\end{aligned}
\end{equation}
The function $\mathfrak{R}$ here is presented in Fig.3.

The final expression of the static transverse suscetibility is
\begin{equation} 
\begin{aligned}
\Pi_{xy}(q)=\frac{igV_{q}^{b}\pi T}{4\pi^{2} S}\int^{\Lambda}_{0}kdk(\frac{1}{\varepsilon_{-}\varepsilon'_{+}}-\frac{1}{\varepsilon_{+}\varepsilon'_{-}}).
\end{aligned}
\end{equation}
By substituting the quasienergy (Eqs.(16)(25)) into the above expression, we obtain
the analytical expression of the term $(\frac{1}{\varepsilon_{-}\varepsilon'_{+}}-\frac{1}{\varepsilon_{+}\varepsilon'_{-}})$
\begin{equation} 
\begin{aligned}
\frac{1}{\varepsilon_{-}\varepsilon'_{+}}-\frac{1}{\varepsilon_{+}\varepsilon'_{-}}=
\frac{1}{(\frac{\sqrt{A+Bk^{2}+k^{4}}}{2m}+\mu)(\mu-\frac{\sqrt{A+B(k+q)^{2}+(k+q)^{4}}}{2m})}
-\frac{1}{(-\frac{\sqrt{A+Bk^{2}+k^{4}}}{2m}+\mu)(\mu+\frac{\sqrt{A+B(k+q)^{2}+(k+q)^{4}}}{2m})},
%1/((mu + (k^4 + B*k^2 + A)^(1/2)/(2*m))*(mu - (A + B*(k + q)^2 + (k + q)^4)^(1/2)/(2*m))) 
%- 1/((mu - (k^4 + B*k^2 + A)^(1/2)/(2*m))*(mu + (A + B*(k + q)^2 + (k + q)^4)^(1/2)/(2*m)))
\end{aligned}
\end{equation}
where we define
\begin{equation} 
\begin{aligned}
A=&4m^{2}(D^{2}+\mu^{2}),\\
B=&4\eta m\mu{\rm cos}2\Phi,\ (\eta=1\ {\rm and}\ \Phi=\pi/2\ {\rm here}).
\end{aligned}
\end{equation}
In order to see the effect of $q$,
we present the result of the term $(\frac{1}{\varepsilon_{-}\varepsilon'_{+}}-\frac{1}{\varepsilon_{+}\varepsilon'_{-}})$ in Fig.4,
with the parameters setted as the usual way (in extrinsic case): $D=0.02$ eV, $\mu=0.2$ eV, and $m=0.268 m_{0}$.
From Fug.4, we can see that,
the maximum value appears as $q=2$,
and
the static transverse susceptibility vanishes at $q=0$ as well as $q\ge 100$.
The orbital diamagnetic susceptibility can be solved by substituting the Eq.(32) into Eq.(24).
We will compare the above results to the isotropic case in below.

While for the linear 2D Dirac system (monolayer),
the static transverse susceptibility at zero temperature reads\cite{Tabert C J,Gorbar E V}
%Pyatkovskiy P K}
\begin{equation} 
\begin{aligned}
\Pi_{xy}(q)=
\frac{-V_{q}^{b}\mu}{ \hbar^{2} v_{F}^{2}}
\left[1-\Theta(q-2k_{F})\left(\frac{\hbar^{2} v_{F}^{2}\sqrt{q^{2}-4k_{F}^{2}}}{2\hbar v_{F}q}-\frac{\hbar^{2}v_{F}^{2}q^{2}-4 m_{D}^{2}}{4\mu \hbar v_{F}q}{\rm arctan}\frac{\hbar v_{F}\sqrt{q^{2}-4k^{2}_{F}}}{2\mu}\right)\right],
\end{aligned}
\end{equation}
and at finite temperature with $q=0$ it reads
\begin{equation} 
\begin{aligned}
\Pi_{xy}(0)=
\frac{-2TV_{q}^{b}}{ \hbar^{2} v_{F}^{2}}
\left[{\rm ln}(2{\rm cosh}\frac{m_{D}+\mu}{T})-\frac{m_{D}}{2T}{\rm tanh}\frac{m_{D}+\mu}{2T}+(\mu\rightarrow -\mu)\right].
\end{aligned}
\end{equation}
Thus the orbital susceptibility can be obtained as
\begin{equation} 
\begin{aligned}
\chi_{{\rm orb}}=\frac{-ge^{2}}{6\pi c^{2}}\lim_{q\rightarrow 0}\frac{-V_{q}^{b}\mu}{2\pi q^{2}},
\end{aligned}
\end{equation}
at zero temperature,
while it vanishes when $q=0$.
At finite temperature,
the non-static (with non-static magnetic field) diamagnetic orbital susceptibility can be obtained (in optical limit) as\cite{Koshino M,Ezawa M2}
%{Anomalous orbital magnetism in Dirac-electron systems Role of pseudospin paramagnetism}
%{Topological phase transition and electrically tunable diamagnetism in silicene}
%{Dynamic current-current susceptibility in three-dimensional Dirac and Weyl semimetals}
%{Theory of orbital magnetism of Bloch electrons: Coulomb interactions}
\begin{equation} 
\begin{aligned}
\chi_{{\rm orb}}^{T}(\Omega)=&\frac{-ge^{2}T\hbar^{2}v_{F}^{2}}{6\pi c^{2}}\sum_{n}\frac{2}{\Omega^{2}+D^{2}}\\
=&\frac{-ge^{2}T}{6\pi c^{2}}\frac{1}{DT}{\rm tanh}\frac{D}{2T}.
\end{aligned}
\end{equation}
The resulting orbital susceptibility has a large peak (anomalous divergence) in the low-energy (or low-temperature) region
as shown in the inset of Fig.7.
While for the parabolic extrinsic metal (with chemicl potential away from the Fermi level) under such weak magnetic field,
the Landau levels are equispaced and won't be affected by the 
magnetic field
except for the ones close to the chemical potential\cite{Thakur A}.
That's distincted from the parabolic Dirac system which with a smaller band touching between the conduction band and valence band
and thus with the non-equispaced Landau levels\cite{Ezawa M3}.

\section{Isotropic dispersion}

Next we analyse the case of isotropic dispersion
where $\phi=atan\frac{k_{y}}{k_{x}}=\pi/4$.
Firstly the eigenenergies in isotropic case can be obtained as
\begin{equation} 
\begin{aligned}
\varepsilon_{\pm}=&\frac{-2m\mu\pm \sqrt{k^{4}+4m^{2}(D^{2}+\mu^{2})+4\eta k^{2}m\mu{\rm cos}2\phi}}{2m}\\
=&\frac{\sqrt{k^{4}+4m^{2}(D^{2}+\mu^{2})}}{2m}-\mu.
\end{aligned}
\end{equation}
The imaginary part of the RPA susceptibility can still be written as
\begin{equation} 
\begin{aligned}
{\rm Im}\ \Pi^{\pm}(\omega,q)=-\frac{g V_{q}^{b}}{4\pi^{2}S}\int^{\Lambda}_{0}dk k
\left(\pi\mp\frac{\pi(-k^{2}k'^{2}-2D^{4}+\varepsilon_{k}^{2}\varepsilon_{k'}^{2})}{\varepsilon_{k}^{2}\varepsilon_{k'}^{2}}\right),
\end{aligned}
\end{equation}
where we consider only the up-spin band in the two-band model.
We follow the parameter setting in above for the extrinsic case,
then the term $\varepsilon_{k}\varepsilon_{k'}$ can be approximately obtained as
\begin{equation} 
\begin{aligned}
\varepsilon_{k}\varepsilon_{k'}\approx & 3k^{4}(1.86\sqrt{(k+q)^{4}+0.0116}-0.2)^2\\
\approx &5.58k^{4}(k+q)^{2},
\end{aligned}
\end{equation}
then the imaginary part of the RPA susceptibility can be obtained 
by solving the above integral.
The result for the isotropic dispersion is presented in the Fig.5.
By comparing the Fig.5 to Fig.1, 
we can see that the imaginary RPA susceptibility in isotropic case is much larger than the anisotropic case,
in other word, the anisotropic dispersion (in parabolic 2D Dirac system) would greatly reduces the polarization.

Next we focus on the important diagonal element of the real space Green's function 
$\int^{\Lambda}_{0}\frac{kdk}{\omega-\varepsilon_{\pm}}$
which is also an important component of the spin susceptibility tensor during the computation of the RKKY interaction
\begin{equation} 
\begin{aligned}
\int^{\Lambda}_{0}&\frac{kdk}{\omega-\varepsilon_{\pm}}\\
=&\mp \frac{1}{\sqrt{A-4m^{2}(\omega+\mu)^{2}}}
\left[\pm m(2m(\omega+\mu){\rm atan}(\frac{k^{2}}{\sqrt{A^{2}-4m^{2}(\omega+\mu)^{2}}})\right.\\
&\left.+2m(\omega+\mu){\rm atan}(\frac{2k^{2}m(\omega+\mu)}{\sqrt{A+k^{4}}\sqrt{A-4m^{2}(\omega+\mu)^{2}}})
+\sqrt{A-4m^{2}(\omega+\mu)^{2}}{\rm ln}(k^{2}+\sqrt{A+k^{4}})
)\right].
\end{aligned}
\end{equation}
The definition of $A$ here is follow the Eq.(34).
Following the parameter setting of the extrinsic case as stated,
we present the above integral in Fig.6 for the static case.
We can see that, for isotropic case,
value of the integral $\int^{\Lambda}_{0}\frac{kdk}{\omega-\varepsilon_{\pm}}$
is smaller that the anisotropic case.
Base on the above result, the analytical expression of the RKKY interaction can be obtained.

To obtain the magnetic susceptibility,
we use the approximated results of the $\varepsilon_{-}\varepsilon'_{+}$ and $\varepsilon_{+}\varepsilon'_{-}$
base on the parameters setted above:
\begin{equation} 
\begin{aligned}
\varepsilon_{-}\varepsilon'_{+}=&2k^{2}(1.86(k+q)^{2}),\\
\varepsilon_{+}\varepsilon'_{-}=&-(2k^{2}-0.4)(1.86(k+q)^{2}).
\end{aligned}
\end{equation}
Then the static transverse suscetibility reads
\begin{equation} 
\begin{aligned}
\Pi_{xy}(q)
=&\frac{igV_{q}^{b}T\pi}{4\pi^{2} S}\int^{\Lambda}_{0}kdk(\frac{1}{\varepsilon_{-}\varepsilon'_{+}}-\frac{1}{\varepsilon_{+}\varepsilon'_{-}})\\
\approx &\frac{igV_{q}^{b}T\pi}{4\pi^{2} S}
\left[
0.268817\frac{1}{q^{2}}(\frac{q}{k+q}-{\rm ln}(k+q)+{\rm ln}k)\bigg|^{\Lambda}_{0}
-\mathfrak{F}
\bigg|^{\Lambda}_{0}
\right],
\end{aligned}
\end{equation}
where we define the function
\begin{equation} 
\begin{aligned}
\mathfrak{F}=&-\frac{1}{(k+q)(-0.2+q^{2})^{2}}\\
&0.24\left[-0.223q+1.118q^{3}+q(k+q){\rm atanh}(2.23k)\right.\\
&\left.+0.559(k+q)(0.2+q^{2}){\rm ln}(0.2-k^{2})-1.118(k+q)(0.2+q^{2}){\rm ln}(k+q)
\right],
\end{aligned}
\end{equation}
After substituting the above expression of static transverse susceptibility $\Pi_{xy}(q)$
into Eq.(24),
we can obtain the final expression of the diamagnetic orbital susceptibility.
Similar to the 3D Weyl semimetal\cite{Thakur A},
the 2D parabolic Dirac system in extrinsic case (with large $\mu$) has positive diamagnetic orbital susceptibility in optical limit.
In Fig.7, we show the static transverse susceptibility and the orbital susceptibility of the isotropic dispersion,
where the anomalous divergent can again be seen.

For an approximated result,
the real space transverse spin susceptibility for parabolic system has obtained by\cite{Kogan E,Stano P}
\begin{equation} 
\begin{aligned}
\Pi_{xy}(r)
\approx \frac{m}{32\pi^{2}\hbar^{2}r^{2}}(1+{\rm cos}(2k_{F}r)),
\end{aligned}
\end{equation}
where $2k_{F}$ here denotes the distance between two Dirac nodes in momentum space.
The oscillation behavior as well as the sublattice dependence will vanishe if we taking average over the unit cell
like the graphene nanotube\cite{Stano P}.
In contrast to the spin susceptibility bilayer (or the massive monolayer) 2D Dirac system which decays as $\sim r^{-2}$,
the monolayer linear Dirac system decays as $\sim r^{-3}$\cite{Brey L},
that implies the parabolic dispersion has a slower decay of the spin susceptibility as the distance increase 
(the distance here $r>\Lambda_{r}$).
After Fourier transformation,
we have
\begin{equation} 
\begin{aligned}
\Pi_{xy}(q)
=&\int d^{2}r\Pi_{xy}(r)e^{-i{\bf q}\cdot{\bf r}}\\
=&\pi I_{0}(iqr)\int^{\Lambda_{r}}_{0}rdr\frac{C}{r^{2}}(1+{\rm cos}(2k_{F}r))\\
\approx &\pi (1-\frac{q^{2}r^{2}}{4})\int^{\Lambda_{r}}_{0}rdr\frac{C}{r^{2}}(1+{\rm cos}(2k_{F}r))\\
=&-\frac{1}{8k_{F}r}\pi C (16k_{F}^{2}r\ {\rm Si}(2k_{F}r)+2k_{F}q^{2}r^{2}+q^{2}r{\rm sin}(2k_{F}r)
+8k_{F}{\rm cos}(2k_{F}r)+8k_{F})\bigg|_{\Lambda_{r}},
\end{aligned}
\end{equation}
where we define, for simplicity, the following function
\begin{equation} 
\begin{aligned}
C=\frac{m}{32\pi^{2}\hbar^{2}},
\end{aligned}
\end{equation}
and ${\rm Si}(z)=\int^{z}_{0}\frac{{\rm sin}x}{x}dx$ is the Sine integral.
The ultraviolet cutoff $\Lambda_{r}$ is related to the range of the double exchange interaction in the kondo lattice\cite{Gulacsi M},
and we here set $\Lambda_{r}=1$ for simplify the calculation.
By subsitituting 
the above transverse spin susceptibility into the Eq.(24),
we obtain the orbital susceptibility as shown in Fig.8,
which is very similar to the one we shown in the inset of Fig.7,
that further confirms the validity of our results.

\section{Summary}

We derive the analytical expressions of the RPA susceptibility (the dynamical polarization),
RKKY Hamiltonian, and the diamagnetic orbital susceptibility in noninteracting case
base on the density-density or current-current correlation function as well as some
important integrals as stated in the text.
Our results obtained in this article are with high precision (for the usual setting of the parameters as
stated in the text)
and valid for the 2D parabolic Dirac systems (no matter with the isotropic or anisotropic dispersion) with a finite gap,
but turn the gap parameter to zero,
our results become valid even for the massless 2D parabolic Dirac systems,
like the bilayer silicene, bilayer graphene, MoS$_{2}$, and the black phosphorus, etc.,
as long as the Rashba-couping and the trigonal warping term are missing.
For the hexagonal Dirac lattice systems,
in virtue of the bipartite feature, we can easily obtain the real space Green's function
and replace its direction degree of freedom to the pseudospin degree of freedom as done in Ref.\cite{Wu C Hrkky},
and it's unaffected by the Rashba-coupling.
That's also in contrast to the Bravais lattice.
%{Phase diagram and magnetic collective excitations of the Hubbard model for graphene sheets and layers}
In bipartite lattice, the RKKY interaction between site impurities is
antiferromagnetic (ferromagnetic) for two sites in oppsite (same) sublattices.
%{RKKY in half-filled bipartite lattices Graphene as an example}
%{Diluted Graphene Antiferromagnet}
Since most of the 2D Dirac system is unlike the QED, which, in non-relativistic case,
the momentum integral is ultraviolet convergent and doesn't need the cutoff,
the most of the 2D Dirac systems as we discussed
need the ultraviolet cutoff $\Lambda$ (i.e., the Fourier transform of the $\Lambda_{r}$),
and we estimate the ultraviolet curoff as the bandwidth of the silicene,
which is about 4.8 eV,
and such estimation is valid and enough for our computation in this article\cite{Sabio J,Saremi S,Gulacsi M}.
%{Magnetism in the dilute Kondo lattice model}
%{Polarization charge distribution in gapped graphene Perturbation theory and exact diagonalization analysis}
While for the effective mass $m$ which is related to both the interlayer and intralayer hopping,
we use the typical value of bilayer silicene which is $m=0.298m_{0}$\cite{Wu C Hele,2,3,4,5},
while for the bilayer graphene, our results are applicable after replace the effective mass as $m=0.033m_{0}$\cite{Sensarma R}
or $m=0.029m_{0}$\cite{Mayorov A S}.
%{Local spin susceptibilities of low-dimensional electron systems}
For the 3D Dirac or Weyl system, the longitudinal susceptibility as well as the current-current correlation function 
is needed to taking the chiral anomaly and the monopole formed by the Weyl nodes into account. 
%{Dynamic current-current susceptibility in three-dimensional Dirac and Weyl semimetals}

\end{large}
\renewcommand\refname{References}

\clearpage

\begin{large}

Fig.1
\begin{figure}[!ht]
   \centering
 \centering
   \begin{center}
     \includegraphics*[width=0.6\linewidth]{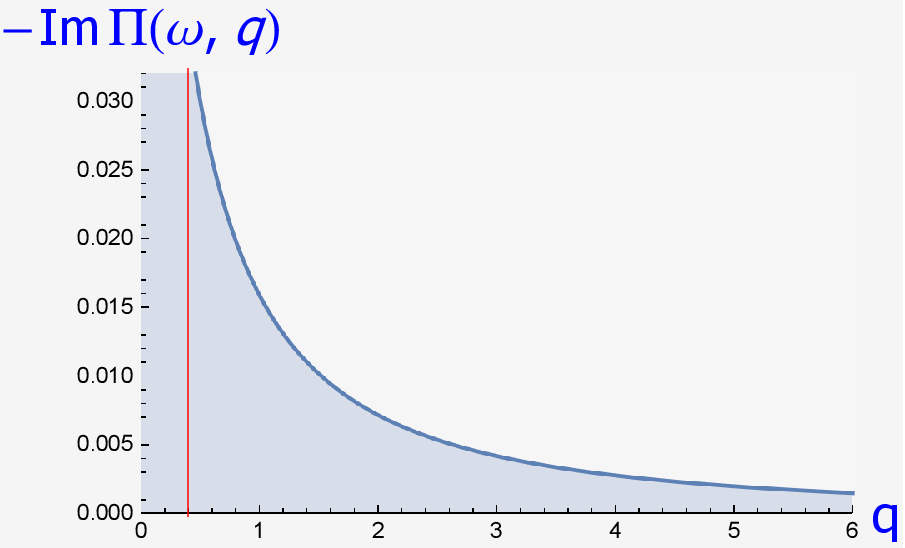}
\caption{(Color online) Imaginary part of the dynamical polarization for frequency $\omega=\varepsilon_{k}-\varepsilon_{k'}$
at zero-temperature. The Dirac-mass is setted as $D=0.02$ eV and the chemical potential is setted as $\mu=0.2$ eV here.
The violet cutoff is setted as $\Lambda=4.8$ eV.
The vertical red line indicates the nesting wave vector $q=2k_{F}=2\sqrt{\mu^{2}-D^{2}}\approx 0.398$.
}
   \end{center}
\end{figure}
\clearpage
Fig.2
\begin{figure}[!ht]
\subfigure{
\begin{minipage}[t]{0.5\textwidth}
\centering
\includegraphics[width=1\linewidth]{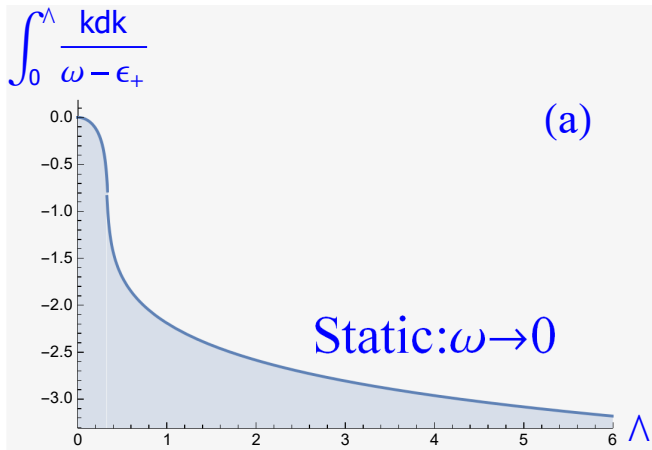}
\label{fig:side:a}
\end{minipage}
}\\
\subfigure{
\begin{minipage}[t]{0.55\textwidth}
\centering
\includegraphics[width=0.9\linewidth]{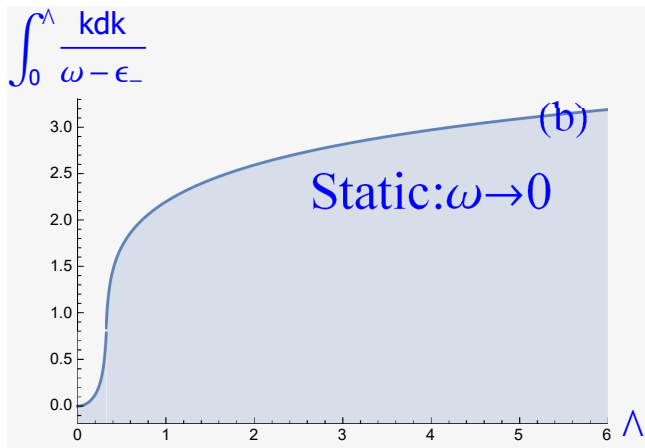}
\label{fig:side:b}
\end{minipage}
}
%{Probing the topological phase transition via density oscillations in silicene and germanene}
\caption{(Color online) The diagonal element $\int^{\Lambda}_{0}\frac{kdk}{\omega-\varepsilon_{\pm}}$ 
(for up-spin band (a) and down-spin band (b))
in static limit $\omega\rightarrow 0$
as a function of the cutoff $\Lambda$ (in unit of eV).
The parameter setting is the same as the Fig.1.
}
\end{figure}
\clearpage

Fig.3
\begin{figure}[!ht]
   \centering
 \centering
   \begin{center}
     \includegraphics*[width=0.6\linewidth]{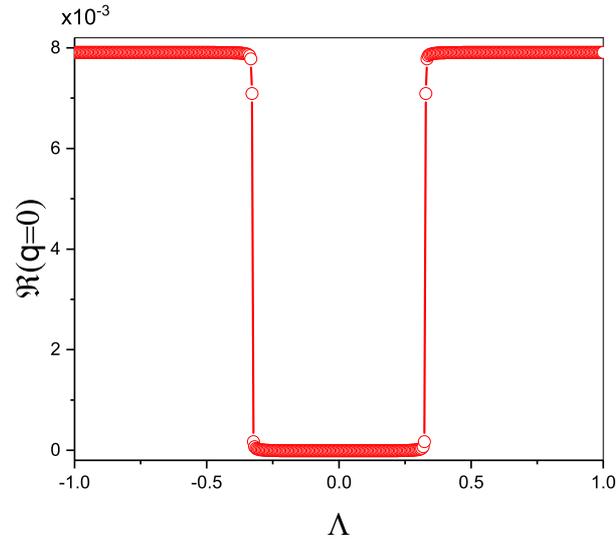} 
\caption{(Color online) Function 
$\mathfrak{R}=\int^{\Lambda}_{0}\frac{kdk}{\varepsilon_{-}\varepsilon_{+}}$ for isotropic dispersion in the case of $q=0$.
}
   \end{center}
\end{figure}

Fig.4
\begin{figure}[!ht]
   \centering
 \centering
   \begin{center}
     \includegraphics*[width=0.6\linewidth]{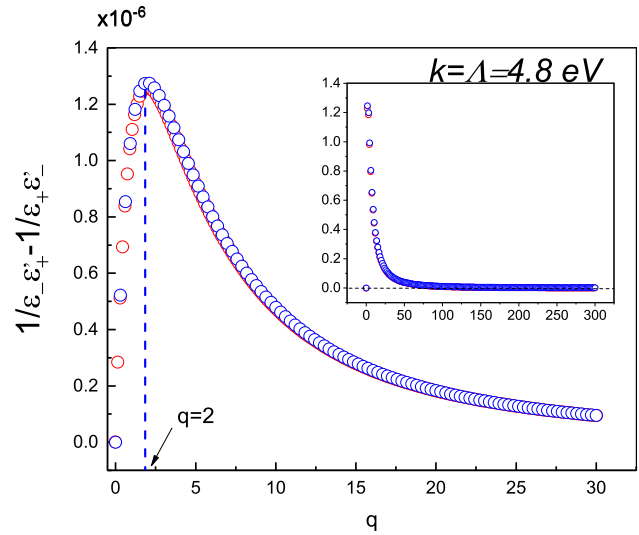}
\caption{(Color online) The term 
$(\frac{1}{\varepsilon_{-}\varepsilon'_{+}}-\frac{1}{\varepsilon_{+}\varepsilon'_{-}})$
as a function of the momentum $q$.
The blue circles correspond to the anisotropic dispersion and the red circles correspond to the
isotropic dispersion.
We can see that the values of the anisotropic case is slightly larger than the ones in isotropic case.
}
   \end{center}
\end{figure}
\end{large}
\clearpage
Fig.5
\begin{figure}[!ht]
   \centering
 \centering
   \begin{center}
     \includegraphics*[width=0.6\linewidth]{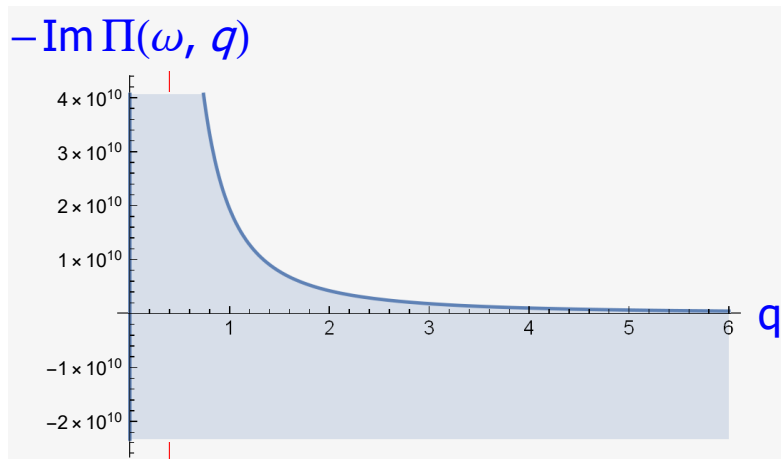}
\caption{(Color online) The same as Fig.1 but for iostropic dispersion.
}
   \end{center}
\end{figure}
\clearpage
Fig.6
\begin{figure}[!ht]
\subfigure{
\begin{minipage}[t]{0.5\textwidth}
\centering
\includegraphics[width=1\linewidth]{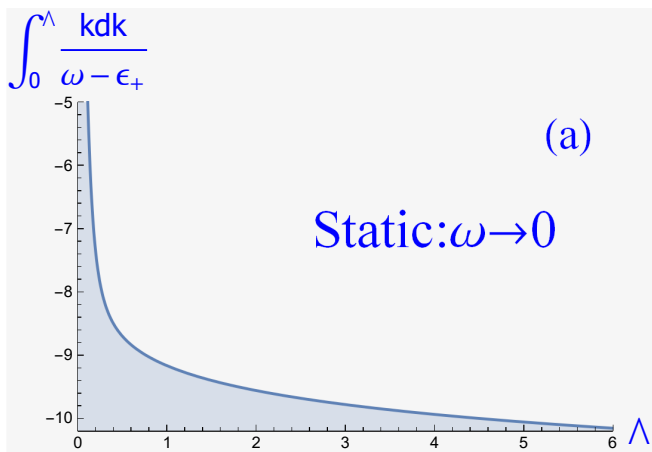}
\label{fig:side:a}
\end{minipage}
}\\
\subfigure{
\begin{minipage}[t]{0.55\textwidth}
\centering
\includegraphics[width=0.9\linewidth]{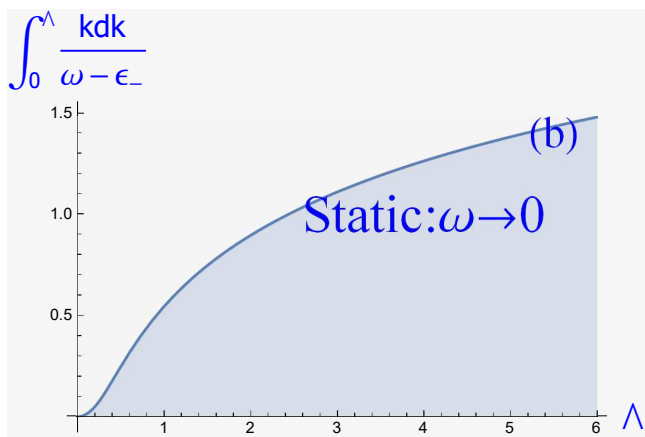}
\label{fig:side:b}
\end{minipage}
}
%{Probing the topological phase transition via density oscillations in silicene and germanene}
\caption{(Color online) The same to the Fig.2 but for the isotropic case.
}
\end{figure}
\clearpage
Fig.7
\begin{figure}[!ht]
   \centering
 \centering
   \begin{center}
     \includegraphics*[width=0.6\linewidth]{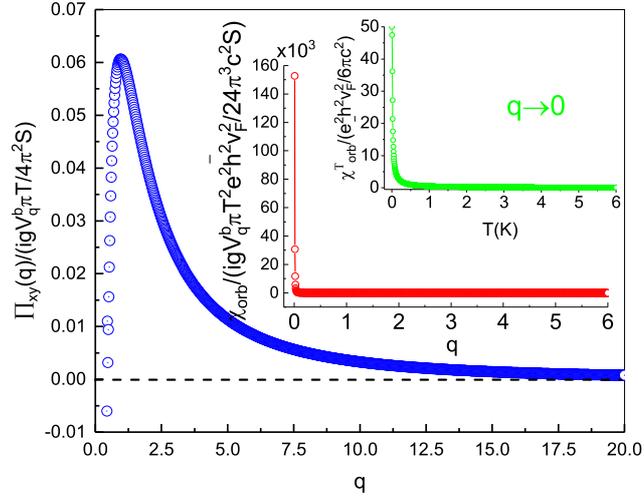}
\caption{(Color online) Static transverse susceptibility (blue circles) and diamagnetic orbital susceptibility 
(red circles) as a function of momentum $q$ at low temperature. The upper inset shows the non-static
orbital susceptibility as a function
of the temperature (green circles).
The large peak of $\chi_{{\rm orb}}$ at low temperature and low momentum can be easily seen.
}
   \end{center}
\end{figure}

Fig.8
\begin{figure}[!ht]
   \centering
 \centering
   \begin{center}
     \includegraphics*[width=0.6\linewidth]{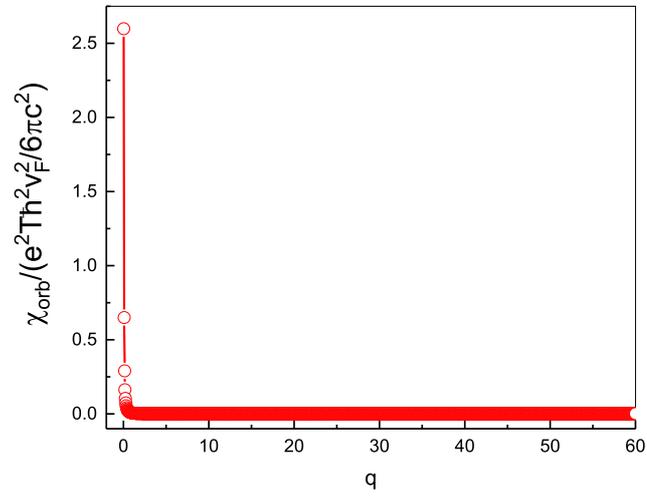}
\caption{(Color online) Diamagnetic orbital susceptibility obtained by Eq.(47).
}
   \end{center}
\end{figure}

\end{document}